\documentclass[a4paper]{jpconf}
\usepackage{graphicx}
\usepackage{txfonts}
\usepackage[square,sort&compress]{natbib}
\bibpunct{[}{]}{,}{n}{,}

\newcommand{\eh}[1]{\,\mathrm{#1}}

\newcommand{\gev}{\eh{GeV}}

\newcommand{\dg}{^{\circ}}

\newcommand{\degr}{$\dg$}

\newcommand{\pct}{\eh{\%}}

\newcommand{\mar}[1]{\mathrm{#1}}

\renewcommand{\epsilon}{\varepsilon}

\newcommand{\tin}[1]{_{\mar{#1}}}

\begin{document}
\title{Phase-resolved Crab pulsar measurements from $25$ to $400\gev$ with the
MAGIC telescopes}

\author{S. Klepser$^1$, G. Giavitto$^1$, M. Lopez$^2$, T.Y. Saito$^3$, T.
Schweizer$^3$, I.~{\v S}nidari\'c$^4$ and R. Zanin$^1$ for the MAGIC collaboration}

\address{$^1$ IFAE, Edifici Cn., Campus UAB, E-08193 Bellaterra, Spain}
\address{$^2$ Universidad Complutense, E-28040 Madrid, Spain}
\address{$^3$ Max-Planck-Institut f\"ur Physik, D-80805 M\"unchen, Germany}
\address{$^4$ Croatian MAGIC Consortium, R. Boskovic Inst., Universities of Rijeka Split, HR-10000 Zagreb, Croatia }

\ead{klepser@ifae.es}

\begin{abstract}
We report on observations of the Crab pulsar with the MAGIC telescopes.
Our data were taken in both monoscopic ($>25\gev$) and stereoscopic
($>50\gev$) observation modes. Two peaks were detected with both modes and
phase-resolved energy spectra were calculated. By comparing 
with Fermi-LAT measurements, we find that the energy spectrum
of the Crab pulsar does not follow a power law with an exponential cutoff,
but has an additional hard component, extending up to at least
$400\gev$. This suggests that the emission above 25 GeV is not dominated by
curvature radiation, as suggested in the standard scenarios of the OG and SG models.
\end{abstract}

\section{Introduction}
The Crab pulsar is the compact object left over after a historic supernova explosion that occurred 
in the year 1054 A.D. 
It is among the brightest known sources at GeV energies.
However, despite numerous efforts, a spectral steepening made
its detection above 10 GeV elusive until 2008. 
Consequently, the presently prefered Outer Gap (OG) \cite{outergap} and Slot Gap (SG)
\cite{slotgap} models predict the emission to be produced through curvature
radiation, which implies an  exponential cutoff at a few GeV. The new satellite-borne
gamma-ray detector Fermi-LAT, which can measure the spectra of gamma-ray
pulsars up to a few tens of GeV, became operational in August 2008.
The spectra measured by Fermi-LAT can be described by a power law with an exponential cutoff,
which supported the OG and the SG model. 

In 2008, the MAGIC-I telescope finally detected the Crab pulsar above 25 GeV
\cite{magicsciencecrab}
with a newly implemented
trigger system, the Sum Trigger \cite{sumtrigger}. Recently, the VERITAS
collaboration reported pulsed gamma rays at energies exceeding $100\gev$
\cite{veritascrab}. However, the cutoff energy of the Crab pulsar spectrum
determined by Fermi-LAT was still $\sim 6$ GeV, which seemed to be in conflict with the
very-high-energy data \cite{fermicrab}.

Here we present the spectral study of the Crab pulsar, using the public Fermi-LAT
data and four years of MAGIC data recorded by the single telescope and
the stereoscopic systems.

\section{The MAGIC telescopes}

The two MAGIC telescopes~\cite{icrcperformance, icrccrabnebula},
situated on the island of La Palma (28.8\degr~N, 17.8\degr~W, $2220\eh{m\,a.s.l.}$), use the Imaging Atmospheric Cherenkov Technique
to detect gamma rays above a few tens of GeV\footnote{The threshold in standard
trigger mode, defined as the peak of the simulated energy distribution for a
Crab-nebula-like spectrum after
all cuts and at low zenith angles, is $75-80\gev$.}. During its monoscopic data
taking era, a low-energy trigger system called Sum Trigger was 
developed and implemented in MAGIC-I in order to search for gamma-ray pulsars. It
reduced the monoscopic energy threshold down to 25 GeV.
%, which resulted in the detection of the Crab pulsar \cite{magicsciencecrab}.

Since MAGIC started operating in stereoscopic mode in summer 2009,
its background suppression was substantially improved and a
sensitivity\footnote{Defined as the source strength needed to achieve $N\tin{ex}/\sqrt{N\tin{bkg}}=5$ in $50\eh{h}$ effective
on-time.}  of
$0.8\pct$ Crab nebula units above $250\gev$ has been achieved
\cite{magicstereoperformance}. 
%The stereo trigger requires a coincidence of the triggers of both telescopes.
Since for technical reasons, the Sum Trigger cannot participate in the stereo
trigger, MAGIC has two independent observation modes for pulsars: Monoscopic,
with lower threshold, and stereoscopic, with lower systematics and better 
sensitivity.
% Therefore, 
%the monoscopic observations have a lower threshold, but the
%stereoscopic observations have significantly lower systematic uncertainties
%due to a better reconstruction and background supression.

\section{Mono-mode observations}

MAGIC-I observed the Crab pulsar with the Sum Trigger
in the winters of 2007/08 and 2008/09.
After a careful data selection, the total effective observation time was 59
hours. 
 
\begin{figure}
   \centering
   $\vcenter{\hbox{\includegraphics[width=7.4cm]{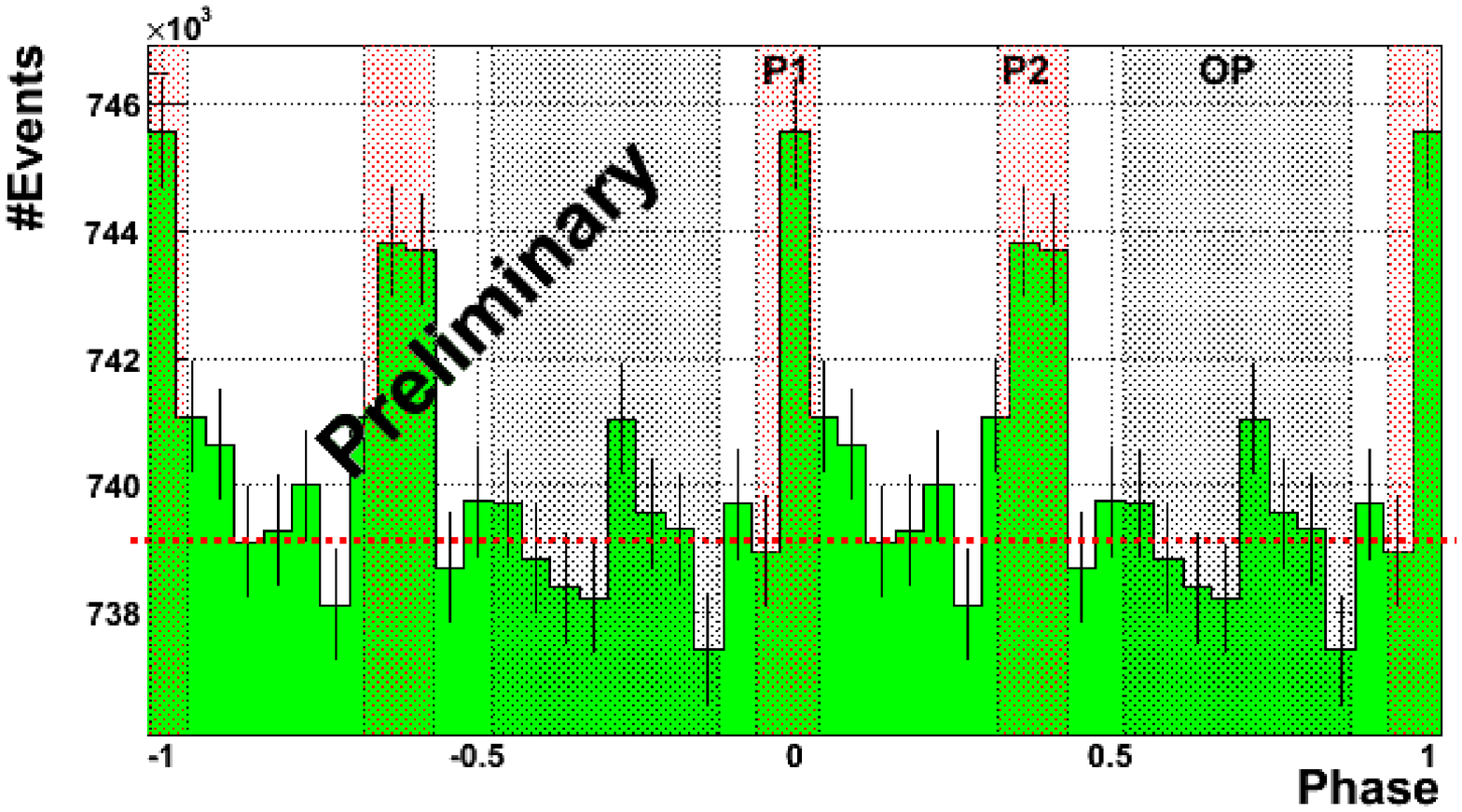}}}$
   \hspace*{.01in}
   $\vcenter{\hbox{\includegraphics[width=7.4cm]{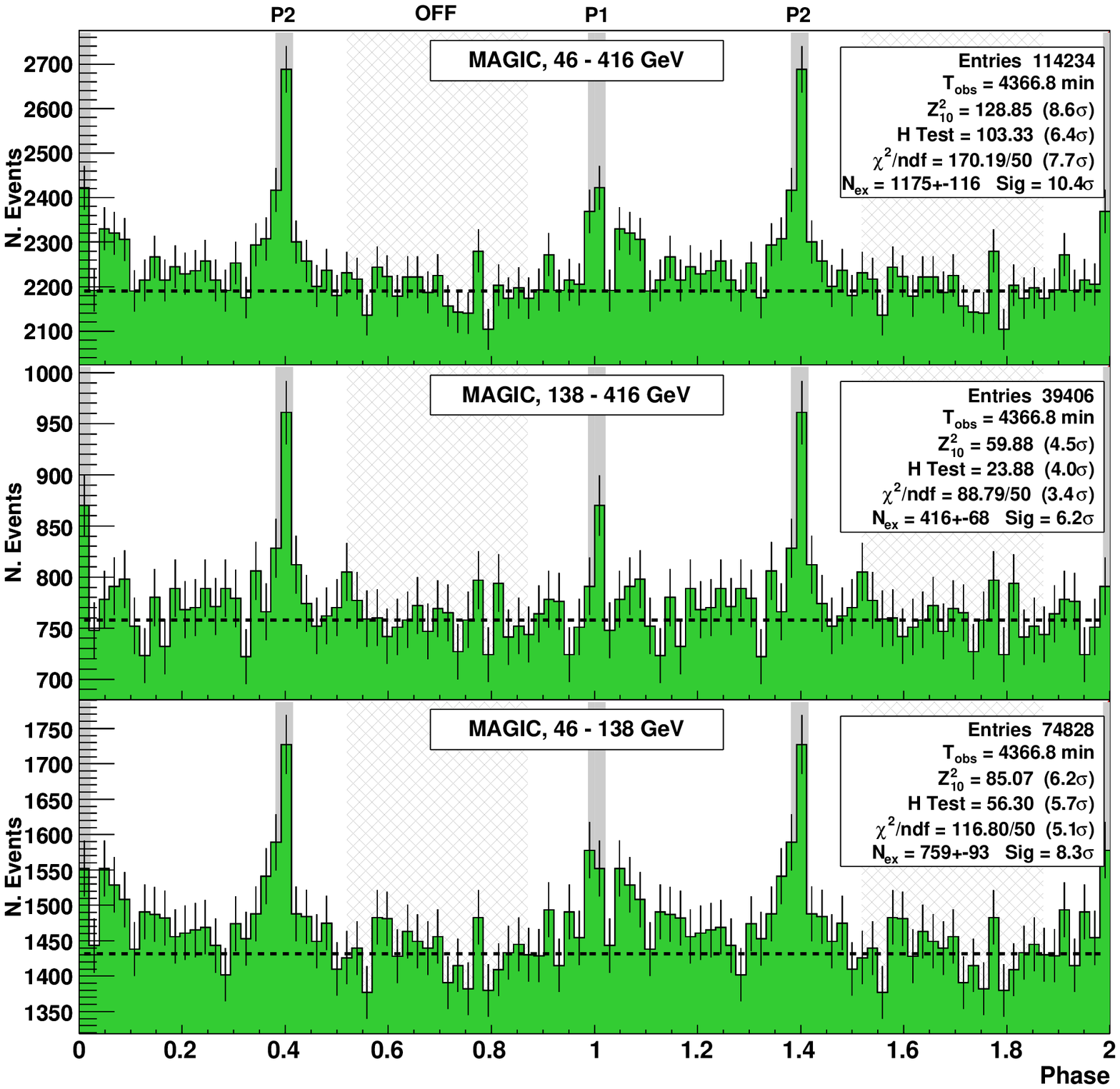}}}$
      \caption{MAGIC folded light curves of the Crab pulsar. Left: Monoscopic
light curve (2007 - 2009). Shaded areas are the on-phase regions
P1/2$_\mar{E}$. Right: Stereoscopic light curve (2009 - 2011) for the total range in
estimated energy, and for two sub-bins separately. The shaded areas are the on-phase regions
P1$_\mar{M}$ and P2$_\mar{M}$ (see text), the light shaded area is the off-region $[0.52 - 0.87]$. The dashed line is the constant background level calculated from that
off-region.
              }
         \label{FigLC}
   \end{figure}

 The light curve of the Crab pulsar obtained with the mono-mode observations is 
shown in Fig. \ref{FigLC} (left). Following the usual convention adopted in
the \textbf{E}GRET era \cite{egretphases} of 
P1$_\mar{E}$ $[0.94 - 0.04]$ and P2$_\mar{E}$ $[0.32 - 0.43]$, the numbers of
excess events in P1$_\mar{E}$ and P2$_\mar{E}$
are $6200 \pm 1400$ (4.3 $\sigma$) and $11300 \pm 1500$ ($7.4\sigma$).
Summing up P1 and P2, the excess corresponds to a significance of 7.5$\sigma$. 
The background level was estimated using the phase interval $[0.52 - 0.88]$.
 
Based on these excess events, the phase-resolved energy spectra of the Crab
pulsar above 25 GeV were computed and are shown as yellow squares in Fig.~\ref{FigSpectra}.
They are compatible with power laws \cite{magictakacrab}.
The energy spectrum measured by Fermi-LAT is also shown in the same figure.
For the Fermi-LAT points, 1 year of the Fermi-LAT data (August 2008 - August 2009) were used.
%The best fit parameters obtained by the unbinned
%likelihood analysis are summarized in Table \ref{TabSpec}.
%The shown spectrum is for the total pulse. 
%Assuming the spectral shape of $F(E)=F_0 E^\Gamma \exp(-E/E_0)$, the likelihood analysis 
%gives $F_0 = (2.32 \pm 0.05) \times 10^{-10}$[cm$^{-2}$s$^{-1}$, MeV$^{-1}$] , $\Gamma = -1.99 \pm 0.02$, $E_c = 6.1 \pm 0.5$ [GeV] as best parameters, which are consistent with the publication 
%by the Fermi-LAT collaboration \cite{FermiCrab}.
The MAGIC-Mono
measurements are not compatible with an extrapolated exponential cutoff
spectrum as determined by Fermi-LAT.
A detailed statistical analysis showed 
the inconsistency amounts to 6.7$\sigma$, 3.0$\sigma$, and 5.8$\sigma$
for (P1~+~P2)$_\mar{E}$, P1$_\mar{E}$ and P2$_\mar{E}$, respectively (see \cite{takathesis} and
\cite{magictakacrab} for more details on the monoscopic analysis).

\section{Stereo-mode observations}

In stereoscopic observations, MAGIC has collected 73 hours of optimal quality Crab
pulsar data between October 2009 and February 2011.
The right panel of Fig.~\ref{FigLC} shows the light curve of the Crab pulsar obtained by 
stereo-mode observations in the total energy range of $46-416\gev$, and two
energy sub-bins of it. 
The significance of the pulsation was tested with the $Z^2_{10}$ test,
the H test \cite{htest}, and a simple $\chi^2$-test. None of these makes an a-priori assumption concerning the position and shape of the
pulsed emission, and they yield significances of $8.6\eh{\sigma}$,
$6.4\eh{\sigma}$ and $7.7\eh{\sigma}$, respectively. By fitting two Gaussians to the two peaks, the peak positions are estimated to be
$0.005 \pm 0.003$ and $0.3996 \pm 0.0014$, while the corresponding FWHMs are $0.025 \pm 0.007$
and $0.026 \pm  0.004$. Defining the signal phases as $\pm 2 \sigma$ of the
fitted Gaussian around the peaks, we calculate spectra both for the
conventional, unbiased P1/2$_\mar{E}$ intervals (see above), and for the
narrower peaks P1$_\mar{M}$ $[0.983 - 0.026]$ and P2$_\mar{M}$ $[0.377 - 0.422]$ that we find in our lightcurve.

The phase-resolved energy spectra are also shown in Fig.~\ref{FigSpectra}.
They connect with the mono-mode measurements within statistical and systematic
errors, and are not incompatible with a power law. 
Further details of the results of the stereo-mode observations 
are presented in \cite{magicstereocrab}.

\section{Discussion and conclusion} \label{discussion}

We found a pulsed VHE gamma-ray signal from the Crab pulsar in monoscopic and
stereoscopic MAGIC data that allows us to present
spectra with an unprecedentedly broad energy range and phase resolution. 
Our spectra range over more than
one order of magnitude, and, along with Fermi-LAT data~\citep{fermicrab}, comprise for the first time a gamma-ray
spectrum of the Crab pulsar from $100\eh{MeV}$ to $400\gev$ without any gap. 
On the high-energy end, they are in good agreement with recently published
VERITAS spectra of P1+P2 above $100\gev$ \cite{veritascrab}, including also the positions and
narrow widths of the two pulses. 

\begin{figure}
   \centering
   $\vcenter{\hbox{\includegraphics[width=5.9cm]{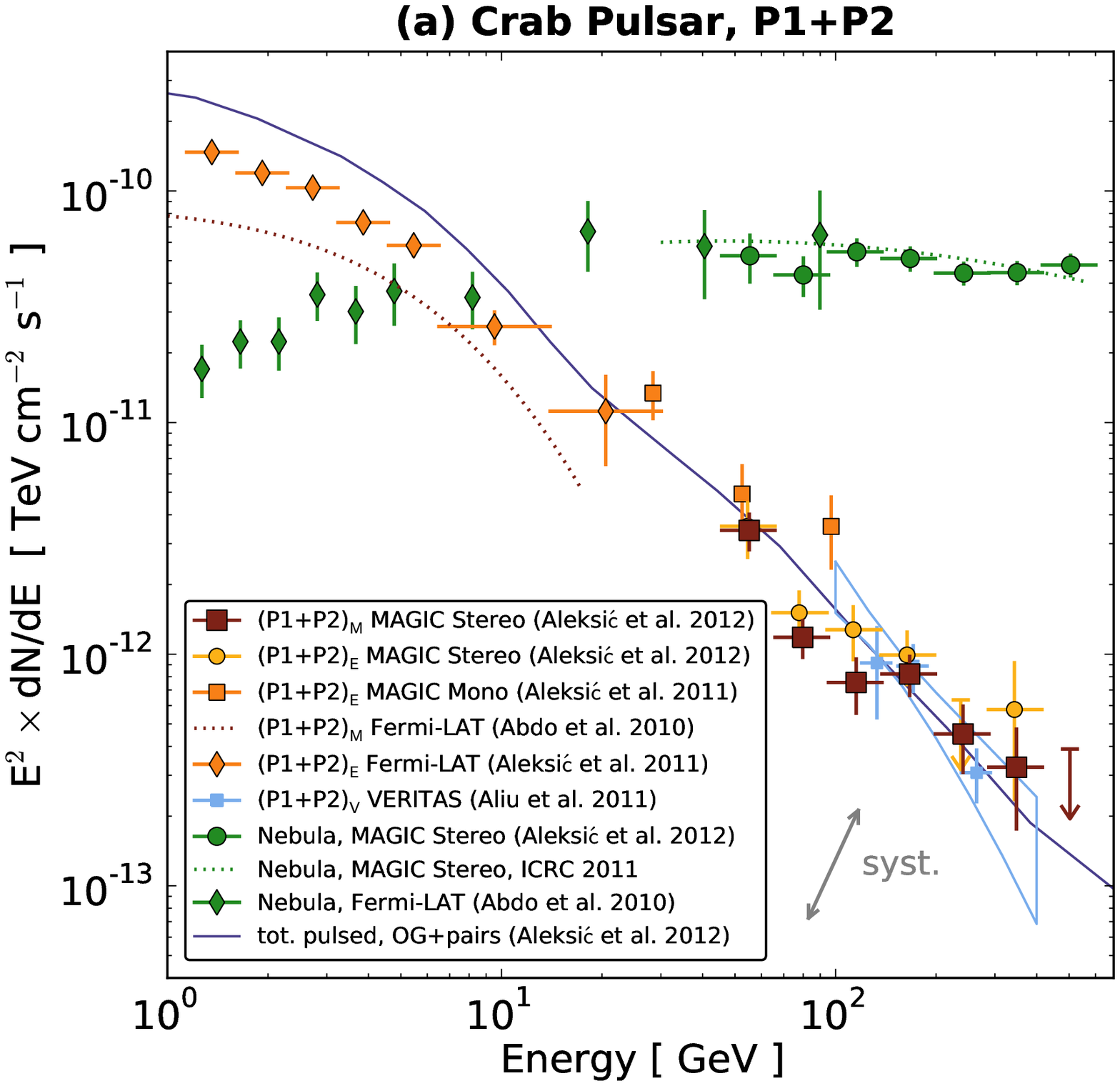}}}$
   $\vcenter{\hbox{\includegraphics[width=4.9cm]{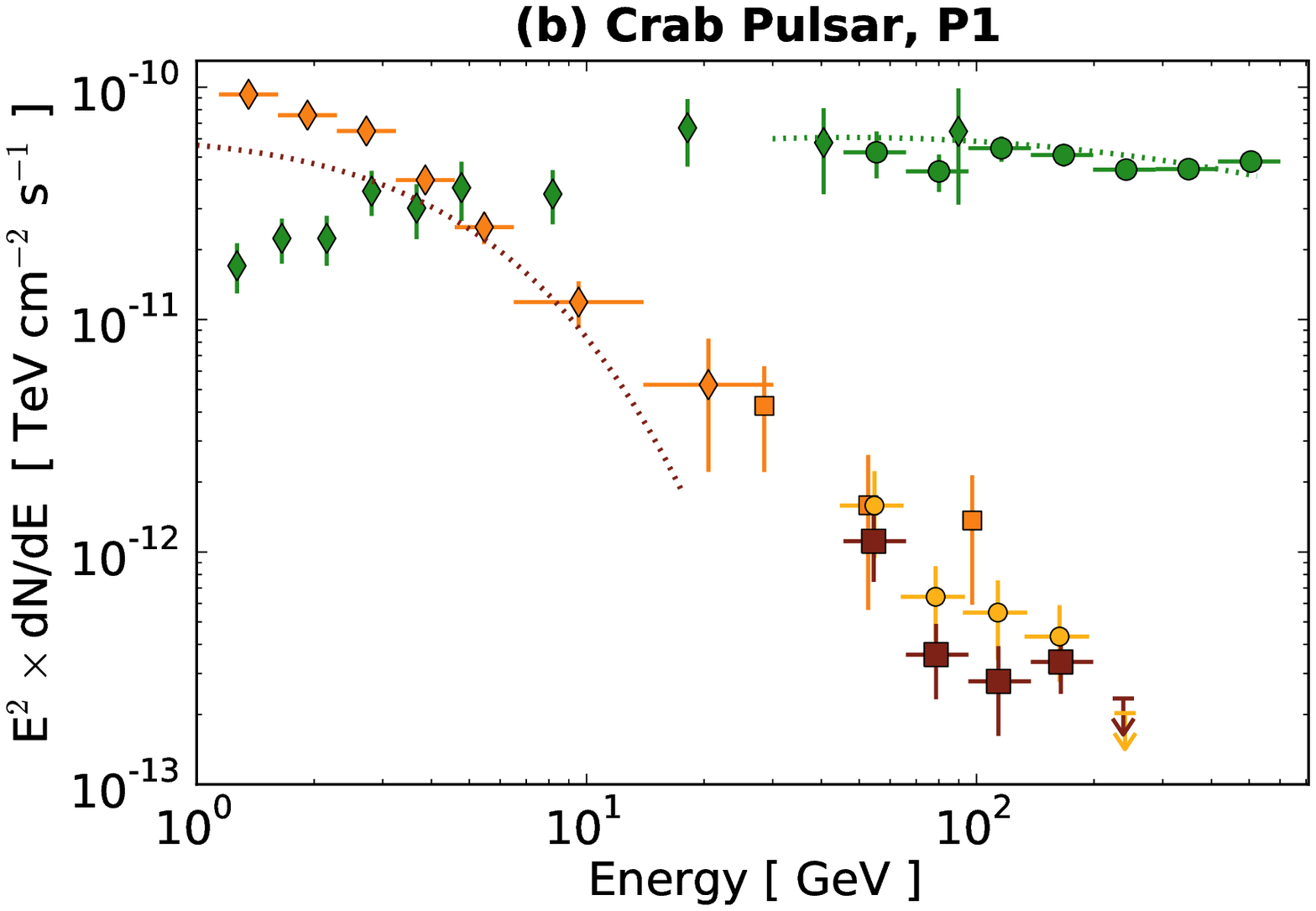}}}$
   $\vcenter{\hbox{\includegraphics[width=4.9cm]{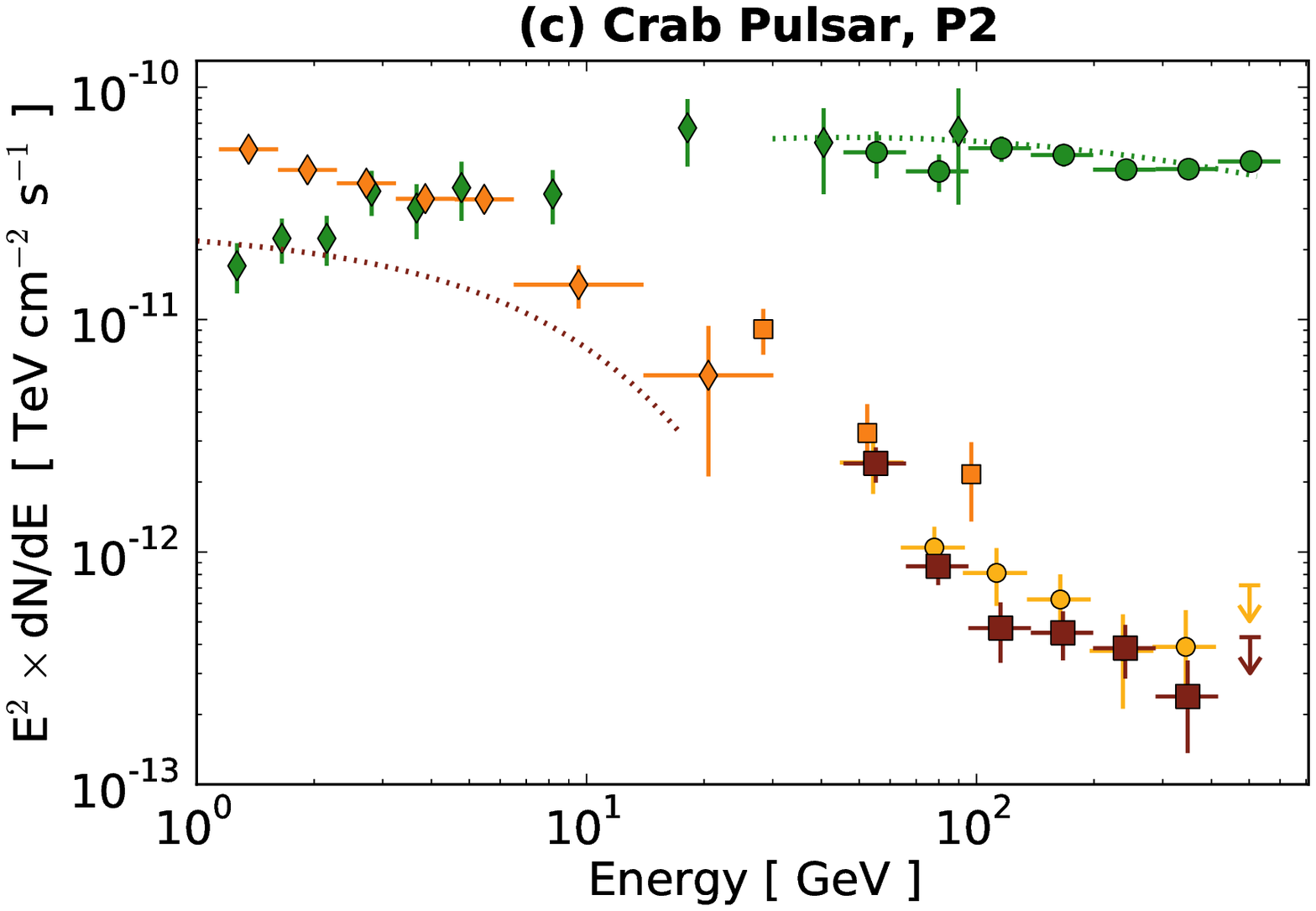}}}$
   \caption{Compilation of spectral measurements of MAGIC and
Fermi-LAT for the two emission peaks P1 and P2 separately, and both
peaks together. The VERITAS spectrum is only available (and shown) for
P1$_{\mar{V}}$+P2$_{\mar{V}}$, which are yet narrower than P1/2$_{\mar{M}}$ For comparison, the Crab nebula
measurements of MAGIC and Fermi-LAT are also shown. Points of similar color refer
to the same phase intervals (marked by the indices \textbf{M}AGIC,
\textbf{E}GRET, \textbf{V}ERITAS, see text). The MAGIC flux points
are bias-corrected using an unfolding technique \cite{unfolding}. The blueish
solid line is the modelling discussed in Section~\ref{discussion}.}
              \label{FigSpectra}%
    \end{figure}

A possible
theoretical explanation of the spectrum~\citep{magictakacrab, magicstereocrab} is that
the spectral tails arise from inverse Compton scattering
of secondary and tertiary electron pairs on magnetospheric IR-UV photons. The
model predicts the onset of that component at a few tens of GeV,
producing a power law spectrum that possibly extends up to TeV
energies. The modeling from~\citep{magicstereocrab} is shown as a blueish solid line in
Fig.~\ref{FigSpectra}(a). It is compatible with the measurements of mono- and stereoscopic
MAGIC, and VERITAS.
It
shall be noted that the GeV flux of the model in~\citep{magicstereocrab} is too
high to match the Fermi-LAT data because it contains bridge emission.
%New calculations are ongoing and indicate
%that the inclusion of processes at higher altitudes might cause an absorption
%of the primary curvature radiation that reduces the flux below a few GeV without
%attenuating the secondary and tertiary VHE components.

%overestimates the has to be compared to the total
%pulsed emission (dash-dotted line), but
%might still overestimate the flux by adopting the vacuum
%rotating dipole solution for the
%magnetic field configuration, which may underestimate the curvature radius. A higher curvature radius, as suggested
%by force-free solutions, might weaken the GeV component, making it consistent
%with the Fermi-LAT data.

%Other possible Ans{\"a}tze include the production of inverse Compton radiation in
%the unshocked pulsar wind outside the light cylinder by pulsed photons
%\citep{aharonianpulsarwind}, or even in a striped pulsar wind
%\citep{stripedwind}.
The two
crucial spectral features to establish in order to benchmark this and other models
are a possible upward-kink in the transition region between curvature and hard
component, and the detection or exclusion of a terminal cutoff at a few hundred
GeV. The MAGIC telescopes, which are being upgraded
in 2011, can provide both of these benchmarks in the coming years, when more
data will improve the statistical precision of the measurements.

\section{Acknowledgments}

We would like to thank the Instituto de Astrof\'{\i}sica de
Canarias for the excellent working conditions at the
Observatorio del Roque de los Muchachos in La Palma.
The support of the German BMBF and MPG, the Italian INFN, 
the Swiss National Fund SNF, and the Spanish MICINN is 
gratefully acknowledged. This work was also supported by 
the Marie Curie program, by the CPAN CSD2007-00042 and MultiDark
CSD2009-00064 projects of the Spanish Consolider-Ingenio 2010
programme, by grant DO02-353 of the Bulgarian NSF, by grant 127740 of 
the Academy of Finland, by the YIP of the Helmholtz Gemeinschaft, 
by the DFG Cluster of Excellence ``Origin and Structure of the 
Universe'', by the DFG Collaborative Research Centers SFB823/C4 and SFB876/C3,
by the Polish MNiSzW grant 745/N-HESS-MAGIC/2010/0 and the Formosa Program between
National Science Council in Taiwan and
Consejo Superior de Investigaciones Cientificas in Spain
administered through grant number NSC100-2923-M-007-001-MY3.

%   \begin{figure}
%   \centering
%   \includegraphics[width=9.0cm]{compilation_P1P2.eps}
%      \caption{Compilation of pulse profile parameters at different energies,
%measured by
%Fermi-LAT~\cite{fermicrab}, MAGIC-Mono~\cite{magictakacrab}, MAGIC-Stereo
%(this work)
%and VERITAS~\cite{veritascrab}. The crosses indicate the phase of the peak,
%while the colored points are the phases of the half-maxima (PHM). The vertical
%lines indicate the phase range
%definitions used for the spectra in
%Fig.~\ref{FigSpectra}.              }
%\label{FigPeaks}
%   \end{figure}

\bibliographystyle{iopart-num}
\bibliography{klepser}

\end{document}